\documentstyle[prd,eqsecnum,aps]{revtex}

\newcommand{\beq}{\begin{equation}} \newcommand{\eeq}{\end{equation}}
\def\bea{\begin{eqnarray}} \def\eea{\end{eqnarray}}
\newcommand{\ba}{\begin{array}} \newcommand{\ea}{\end{array}}

\def\a{\alpha} \def\b{\beta} \def\g{\gamma} \def\G{\Gamma}
\def\d{\delta} \def\D{\Delta}  
   
 \def\l{\lambda}  \def\m{\mu} \def\n{\nu}
  \def\p{\pi}  \def\r{\rho}
  \def\t{\tau} 
 \def\f{\phi}   
   
\def\pa{\partial}  
\def\sqr#1#2{{\vcenter{\hrule height.#2pt \hbox{\vrule width.#2pt
height#1pt \kern#1pt \vrule width.#2pt} \hrule height.#2pt}}}
\def\sq{{\mathchoice\sqr55\sqr55\sqr{2.1}3\sqr{1.5}3}\hskip 1.5pt}
\def\fletxeta{{\buildrel \leftrightarrow \over \partial}}
\def\IR{{I\kern-0.25em R}}

\begin{document}

 \draft

\title{ A Generic Renormalization Method in Curved Spaces and at Finite
Temperature\footnote{Research supported in part by CICYT grant
\#AEN93-0695, and NATO grant \#910890.\\ E-mail: {\tt
comellas@ebubecm1, hagensen@ebubecm1, latorre@ebubecm1}.}}
\author{Jordi Comellas, Peter E. Haagensen and Jos\'e I. Latorre}
\address{Departament d'Estructura i Constituents de la Mat\`eria,
\\Facultat de F\'\i sica, Universitat de Barcelona\\ Diagonal, 647\ \
08028 Barcelona, SPAIN}

\maketitle

\begin{abstract}
Based only on simple principles of renormalization in coordinate space,
we derive closed renormalized amplitudes and renormalization group
constants at 1- and 2-loop orders for scalar field theories in
general backgrounds.  This is achieved through a generic renormalization
procedure we develop exploiting the central idea behind differential
renormalization, which needs as only inputs the propagator and the
appropriate laplacian for the backgrounds in question.  We work out this
generic coordinate space renormalization in some detail, and
subsequently back it up with specific calculations for scalar theories
both on curved backgrounds, manifestly preserving diffeomorphism
invariance, and at finite temperature.
\end{abstract}
\vspace{1cm}
\pacs{UB-ECM-PF 94/10}

\setcounter{page}{1}

\section{ Generic renormalization of short-distance singularities}
\bigskip

Quantum field theories are often considered under classical external
backgrounds.  Two standard examples are thermal baths \cite{kapusta} and
curved space-times \cite{birrell} and, in general, the procedure of
renormalization becomes much more involved due to the appearance of new
dimensionful scales in the problem.

It is nevertheless known that some of the first coefficients of the
renormalization group constants remain unaltered.  For instance, finite
temperature can be encoded by compactifying the Euclidean time, and this
obviously affects the long-distance properties of the theory but not the
leading short-distance ones.  It is natural to expect no modification of
the first coefficient of the $\b$-function.  Similarly, gravitational
backgrounds do not affect the leading renormalization of flat field
theories\footnote{ But there will be, of course, leading singularities
associated to the renormalization of new couplings like $\xi R \phi^2$,
etc..} due to the Equivalence Principle, which allows for a locally flat
space-time and thus guarantees that short-distance singularities are
those of Minkowski space-time.  On the other hand, the complicated
structure of thermal and gravitational backgrounds makes it often
impossible in practice to get closed expressions for even the simplest
Green functions.  Infrared modifications of a theory do not interfere
with the renormalization of leading singularities but make almost
impractical any computation at and beyond one loop.

It is our aim to present a renormalization procedure adapted to quantum
field theory defined at finite temperature or in curved space-times
which produces closed explicit expressions for Green functions at low
orders of perturbation theory.  The method is based on the ideas behind
differential renormalization (DR hereafter)\cite{fjl}, and we shall now
expose it in a general setting.  By way of example, we shall also check
that the first two coefficients of the $\b$-function and the first one
of the anomalous dimension of the scalar field in $\l\f^4$ theory are
independent of long-distance aspects of the theory.

Let us consider a massless Euclidean  scalar field theory with a
propagator given by
\beq{<\f (x)\f (0)>\ \ \equiv\ \ G(x,0;\alpha)\ ,\label{prop}}\eeq
where $\alpha$ stands for dimensionful parameters such as temperature
or curvature. The propagator obeys an equation of the type
\beq{\cal D} \ G(x,0;\alpha) = {\delta^{(4)}(x) \over
\sqrt{g(x)}}\,,\label{laplacian}\eeq
where ${\cal D}$ is the proper scalar laplacian of the theory, for
instance on curved space-time (with background metric $g_{\mu\nu}(x)$)
or with nontrivial boundary conditions when necessary.  For a preview
of a particular form of this operator, one may consider for example
\beq{{\cal D}=-g^{\m\n}(x)\nabla_\m\pa_\n -\a^2\ ,
\label{previewlap}}\eeq
where $\nabla_\mu$ stands for a covariant derivative and $\a^2$ is
related to the curvature of the background metric (see Sec. 2).  For the
sake of simplicity, we shall first consider massless $\lambda \phi^4$
theory.
The massive case is discussed at the end of this section.  We shall
assume that our field theory is locally flat, that is,
\beq{G(x,0;\alpha)\ \ {\buildrel x\to 0 \over \longrightarrow } \ \
{1\over 4\pi^2} {1\over x^2} \label{shortlimit}}\eeq
This is the case of field theories both in curved spaces or in thermal
baths.  The long-distance behavior of the propagator will surely be
complicated.  This is at the heart of the problem of defining a Fourier
transform \cite{bunch,vil} and computing it.
\begin{figure}

\protect\[ 
  \begin{picture}(100,45)(0,7)
    \thicklines
    \put(30,10){\line(-1,-1){10}}
    \put(30,10){\line(-1,1){10}}
    \put(50,10){\circle{40}}
    \put(70,10){\line(1,1){10}}
    \put(70,10){\line(1,-1){10}}   

  \end{picture}
\protect\] 

\end{figure}
\begin{center}{\it Figure 1.}\end{center}\medskip

As we turn to the simplest quantum correction of the amputated
four-point vertex function, Fig. 1, we find the  contribution
\beq {\l^2\over 2}\ (G(x,0;\a ))^2\ ,\label{bubble}\eeq
which displays a leading singularity at short distances of the type
\beq{\left(G(x,0;\alpha)\right)^2\ \ {\buildrel x\to 0 \over
\longrightarrow}\ \ {1\over 16 \pi^4} {1\over (x^2)^2}+\ldots\
.\label{shortdistance}} \eeq
Due to this (logarithmic) singularity at $x=0$, Eq.  (\ref{bubble}) does
not accept a Fourier transform, it is not a good distribution upon
integration against plane waves and needs to be renormalized.
Typically, the short-distance expansion in Eq.  (\ref{shortdistance})
will contain subleading terms which diverge as the propagator itself and
need not undergo renormalization.  In the absence of nontrivial
backgrounds, the method of differential renormalization gives a recipe
to obtain right away the renormalized version of Eq.  (\ref{bubble}).
The idea is to rewrite this equation by extracting a laplacian operator,
\beq{ {1\over (x^2)^2} = -{1\over 4}\ \sq\ {\ln x^2 M^2\over
x^2}\,,\label{boxtrick}}\eeq
where $\sq\equiv \partial_\mu\partial_\mu$.  This expression is an
identity at $x\not= 0$ and, furthermore, produces an extension of the
too singular function of the l.h.s. into a proper distribution on the
r.h.s., provided the operator $\sq$ is understood as acting onto the
left.  We should now generalize this simple identity to more complicated
cases, when the infrared behavior of the propagator becomes very
involved.  We achieve this goal by noting that, in general, the
logarithmic singularity in Eq.  (\ref{bubble}) is solved by writing it
as
\beq{ G^2(x) =- {1\over 16 \pi^2}\, {\cal D} \left( G(x) \ln
G(x)/M^2\right)+ D(x)\,.\label{genbox}}\eeq
$D(x)\equiv D(x,0;\alpha)$ is then a {\it bona fide} distribution which
will become explicit in each particular case, and $G(x)$ is short-hand
for $G(x,0;\alpha)$.  Again, the function on the l.h.s. is turned into a
distribution while nothing is changed away from the singularity.  It is
easy to see, by using the leading behavior of the propagator, that
indeed the above formula encodes the renormalization of the singularity
as shown in Eq.  (\ref{boxtrick}).  The form of $D(x)$ will depend on
the particular problem we are addressing, and may look involved since it
is encoding long-distance properties of the theory, but it is guaranteed
to be at most as singular as the propagator itself.

There are basically three reasons why Eq.  (\ref{genbox}) produces the
correct renormalization of the 1-loop diagram in thermal or
gravitational deformations of massless $\l\f^4$ theory:

1) Removal of singularities.  As we just explained, our generalized
renormalization of the bubble diagram is guaranteed by the
short-distance limit of all the pieces of the r.h.s. of Eq.
(\ref{genbox}).

2) Renormalization group behavior.  The renormalization scale only
appears within the logarithm.  The renormalization group equation
at one loop\footnote{New $\beta$-functions associated to mass and
background couplings will eventually appear, but at higher loop order.}
\beq{\left( M{\pa \over \partial M}+\beta (\lambda ){\partial \over
\pa\l } -4 \g (\l )\right) \Gamma^{(4)}=0\ , }\eeq
where $\g (\l )$ stands for the anomalous dimension of the $\phi$ field,
$\beta(\lambda)$ is the $\b$-function and $\Gamma^{(4)}$ is the
amputated four-point function, will thus contain the piece
\beq{M{\pa\over\pa M}\left(-{1\over 16\pi^2}\,{\cal D}\left( G(x)\ln
G(x)/M^2\right)+D(x)\right)={1\over 8\pi^2}{\cal D}G(x)={1\over 8\pi^2}
{\d^{(4)}(x)\over\sqrt{g(x)}}\,.}\eeq
The full diagram carries an extra $\lambda^2/2$ factor plus the addition
of the $s,t$ and $u$ channels.  Therefore, the $M$ dependence can be
reabsorbed by a change of the tree-level coupling dictated by the beta
function \beq{\beta(\lambda)=M{\pa\l\over\pa M} ={3 \lambda^2\over 16
\pi^2}\,,\label{betafunction}}\eeq and the anomalous dimension receives
no contribution from this diagram.

3) Unitarity.  The imaginary part of the amplitude is related to the
cross-section through the standard unitarity relations.  In Eq.
(\ref{genbox}), the imaginary part is carried by the logarithm, which in
turn gives rise to a delta function.  As expected, the imaginary part of
the 1-loop contribution is then proportional to the tree level
structure.

It is also noteworthy that in the presence of a gravitational
background, our procedure preserves diffeomorphism invariance since only
covariant derivatives are manipulated.  There is just the desired scale
invariance breakdown.  On the other hand, the above recipe has to be
slightly modified in the case of massive theories as we discuss at the
end of this section.
\begin{figure}
\protect\[ 
  \begin{picture}(100,45)(0,7)
    \thicklines
    \put(30,10){\line(-1,1){10}}
    \put(30,10){\line(-1,-1){10}}
    \put(50,10){\circle{40}}
    \put(90,10){\circle{40}}
    \put(110,10){\line(1,1){10}}
    \put(110,10){\line(1,-1){10}}   

  \end{picture}
\protect\] 
\end{figure}
\begin{center}{\phantom{xxxx}\it Figure 2.}\end{center}
\begin{figure}
\protect\[ 
  \begin{picture}(100,45)(0,7)
    \thicklines
    \put(30,10){\line(-1,1){10}}
    \put(30,10){\line(-1,-1){10}}
    \put(60,-07){\line(0,1){35}}
    \put(50,10){\circle{40}}
    \put(60,-07){\line(2,-1){10}}
    \put(60,28){\line(2,1){10}}

  \end{picture}
\protect\] 
\end{figure}
\begin{center}{\phantom{xxxx}\it Figure 3.}\end{center}
\bigskip

At two loops, the amputated 4-point function is given by the two
diagrams in Figs. 2,3.  The first diagram corresponds to a convolution
and, thus, introduces no new kind of singularity.  It just produces a
promotion of leading logs to second order.  To see this, note that the
bare amplitude \beq{\int d^4 w \ \sqrt{g(w)}
\left(G(x,w;\alpha)\right)^2 \left(G(w,0;\alpha)\right)^2
\label{bareconvolution}}\eeq simply gets renormalized using Eq.
(\ref{genbox}).  The fact that this diagram brings no contribution to
the 2-loop $\b$-function stems from the fact that its renormalization
scale dependence is non-local, which is readily checked \bea
{\displaystyle M{\pa\over\pa M}\ \left[ \ \int d^4w \ \sqrt{g}\right.}&
{\displaystyle \left(-{1\over 16\pi^2}\, {\cal D} (G(x,w;\alpha) \ln
G(x,w;\alpha)/M^2 )+D(x,w;\alpha)\right)} \cr \times
&{\displaystyle\left. \left(-{1\over 16\pi^2}\, {\cal D} (G(w,0;\alpha)
\ln G(w,0;\alpha)/M^2 )+D(w,0;\alpha)\right)\right]}\cr {\displaystyle
={1\over 4\pi^2} }&{\displaystyle \left(-{1\over 16\pi^2}\,{\cal
D}(G(x,0;\alpha) \ln G(x,0;\alpha)/M^2) +D(x,0;\alpha)\right) \
.}\label{renconvolution}\eea

On the other hand, the diagram in Fig. 3 displays a new kind of
singularity, a generic 3-point one.  The bare form of the amplitude
reads
\beq{G(x,0;\alpha)\, G(y,0;\alpha) \left(G(x,y;\alpha)\right)^2\
.\label{bareform}}\eeq
Let us recall \cite{fjl} that ordinary flat space renormalization goes
in two steps, first renormalizing the inner 2-point singularity
\beq{{1\over x^2}{1\over y^2}{1\over (x-y)^4}=
-{1\over 4} {1\over x^2}{1\over y^2}\ \sq\ {\ln (x-y)^2
M^2\over (x-y)^2}\,.\label{cone}}\eeq
Then, pulling derivatives out with exact manipulations,
one finds
\beq{-{1\over 4}
{\partial\over \partial y_\mu}\left({1\over x^2 y^2}\,
{ \fletxeta  \over \partial y_\mu}
{\ln(x-y)^2M^2\over
(x-y)^2}\right)+\pi^2\delta^{(4)}(y)\,\, {\ln x^2M^2\over x^4}\,. }\eeq
This final piece needs the following supplementary, genuine 2-loop
renormalization
\beq{{\ln x^2M^2\over x^4}= -{1\over 8}\ \sq\
{\ln^2 x^2M^2 +2 \ln x^2\mu^2\over x^2}\,.\label{newtrick}}\eeq
The coefficient in front of this last expression is responsible for
yielding the correct 2-loop $\b$-function,
\beq\label{betafunct}\beta(\lambda)= {3 \lambda^2\over 16 \pi^2}
-{17\over 3}{\lambda^3\over 256\pi^4}\ ,\eeq
once the first term of the anomalous dimension of $\phi$ is introduced
in the renormalization group equation.
Coming back to our general discussion, it is easy to guess the correct
generic renormalization of massless deformations of the flat theory.
The first step is similar to the one in Eq. (\ref{cone}),
\beq\label{firststep}G(x,0;\alpha)\, G(y,0;\alpha)\left( -{1\over
16\pi^2 }\,{\cal D}\left(G(x,y;\alpha) \ln G(x,y;\alpha)/M^2\right)
+D(x,y;\alpha)\right)\ .\eeq
Through exact manipulations of covariant derivatives (recall that
typically, ${\cal D}= -g^{\mu\nu} \nabla_\mu\pa_\nu-\a^2$, where $\a^2$
is a constant), the expression will be transformed into a total
derivative plus a generic 2-loop singularity which will be
renormalized by
\beq\label{generictwoloop}G^2(x)\ln G(x)/M^2 = -{1\over 32 \pi^2}\,{\cal
D}\ \left[\ G(x)\left(\ln^2 {G(x)\over M^2}-2 \ln {G(x)\over
M^2}\right)\right]+D'(x;M)\, .\eeq
$D'(x;M)$ is again some left-over distribution. By taking the
logarithmic $M$ derivative of the above, we find
\beq\label{mddm}M{\pa\over\pa M}D'(x;M)=-2D(x)\ ,\eeq
where $D(x)$ is the distribution appearing in Eq.
(\ref{genbox}). This integrates to
\beq\label{dprime}D'(x;M)=D(x)\ln G(x)/M^2+f(x)\ ,\eeq
where $f(x)$ is some distribution independent of $M$.  It turns out that
this dependence is precisely what is needed for consistency of the RG
equations for $\G^{(4)}$ to order $\l^3$.  This is another instance of
the generality of our procedure.  Both $D(x)$ and $f(x)$ give rise to
benign singularities of the type
\beq\label{singprecision}{1\over x^2}{1\over y^2}{1\over (x-y)^2}\ ,\eeq
which are integrable.  Another fine point in Eq.  (\ref{generictwoloop})
is that the relative coefficient $-2$ between the two powers of log is
fixed by the identity itself and eventually produces the same 2-loop
$\b$-function as in the flat case.  The fact that we have renormalized
logarithmic singularities has left no room for subleading divergences
which might depend on $\alpha$, leaving the result for the $\b$-function
unaltered.  It is, indeed, remarkable how universal these manipulations
are.  They substantiate the common lore about independence of some
short-distance singularities on long-distance physics, while keeping
closed explicit expressions for the whole amplitudes.\bigskip

\begin{figure}

\protect\[ 
  \begin{picture}(100,45)(0,7)
    \thicklines
    \put(20,10){\line(1,0){10}}
    \put(50,10){\circle{40}}
    \put(30,10){\line(1,0){40}}
    \put(70,10){\line(1,0){10}}   

  \end{picture}
\protect\] 
\vspace{1ex}
\end{figure}
\begin{center}{\it Figure 4.}\end{center}
\bigskip

Let us add a final and nontrivial example of this generic
renormalization.  The first non-local wave-function renormalization can
again be treated in full generality.  The relevant diagram is shown in
Fig. 4 and we shall call it the ``setting sun".  Its generic
renormalization is
\beq\label{settingsun}G^3(x)={1\over 512 \pi^4}\, {\cal D}\,{(\cal
D}+b)\, G(x) \ln G(x)/M^2 + D''(x)\,.\eeq
where $b$ is a constant that depends on the details of the problem.  It
accompanies the term with a single ${\cal D}$ operator, which is needed
to cure subleading divergences which, for the first time, may introduce
temperature and curvature dependences.  As in all previous cases,
$D''(x)$ is a left-over distribution which carries information on the
long-distance properties of the theory but does not affect
short-distance renormalizations.  A small effort applying the
renormalization group equation yields the result for the
leading anomalous dimension of the field $\phi$
\beq\label{anomalous}\g (\l )={1\over 12}{\left(\lambda\over 16
\pi^2\right)^2} \ .\eeq
This generic renormalization again fulfills all three properties listed
above: removal of singularities, correct renormalization group behavior
and unitarity.

Let us briefly discuss our renormalization procedure in the presence of
masses. First of all, the equation for the free propagator
changes from Eq. (\ref{laplacian}) to
\beq\label{masslap}({\cal D}+m^2) G(x,0;\alpha;m)={\delta^{(4)}(x)
\over \sqrt{g(x)}}\,.\eeq

{}From this equation and the general form of the laplacian as previewed in
Eq.  (\ref{previewlap}), it becomes clear that the mass and the $\a$
parameter play similar roles.  Their presence does not interfere with
the leading singularities of the theory\cite{massive}.  They first
enter actively the game of renormalization to control the subleading
renormalization of the 2-point function, which we called $b$ in Eq.
(\ref{settingsun}).  Nonetheless, a blind application of massless
recipes leads to unnecessarily complicated renormalized expressions due
to the following observation.  At long distances, the correct behavior
of, say, the bubble diagram will be an exponential fall off as $\exp(-2
m r)$ since two particles circulate through the loop diagram.  This is
not the behavior of ${\cal D}( G \log G/M^2)$, which decays as $\exp(- m
r)$.  Were we to use Eq.  (\ref{genbox}), the left-over distribution $D$
would have to correct for this long-distance behavior.  Although
renormalization would be carried out correctly, the answer achieved
would not be minimally representing the long-distance fall-off.  The
solution to the problem is to substitute $\log G/M^2 $ by a function
which shares the same short-distance behaviour but provides a better
representation of long-distance physics.  A first step in this direction
is to notice through examples that the more appropriate differential
operator to be used to renormalize the bubble diagram turns out to be
${\cal D} + 4 m^2$.  A production threshold naturally finds its place
next to the laplacian (and eventually sneaks in in momentum space as
$p^2+4m^2$), although it multiplies a distribution and thus there was
no absolute need for it.  The change in the distribution following the
operator is of a more refined nature, and we pospone its discussion to
the next section after an explicit example.  Let us mention here that
unitarity by itself demands further changes besides the one in the
differential operator.

We have just sketched a program to get explicit Green functions at low
orders of perturbation theory for $\lambda\phi^4$ on modified spaces
which leave the leading behavior of the propagator unchanged.  The
universality of the first two coefficients of the $\b$-function and the
first one in the anomalous dimension of the field $\phi$ also follows
from the above.  We shall now consider the explicit realization of these
ideas.  In Sec. 2, we treat constant curvature backgrounds, where we can
in fact develop DR in an exact way and maintain manifest coordinate
invariance, and in Sec. 3 we consider finite temperatures.

\section{renormalization in maximally symmetric spaces}\bigskip

The renormalization effort in curved spaces, to date, has centered
mainly on the techniques of point-splitting \cite{christensen},
background field and heat kernel methods \cite{birrell,toms}.  As
opposed to flat space, in curved manifolds coordinate- rather than
momentum-space is the natural setting for calculations, due to the lack
of a generically-defined Fourier transform \cite{bunch,vil}.
Because of that, it is reasonable to expect that differential
renormalization will also be a natural method, beyond the abovementioned
ones, since it is also based on coordinate space.  The hope is that DR
will present itself as an efficient renormalization procedure in curved
space, and as we will see, this hope is brought to fruition.  We also
find that DR manifestly preserves the diffeomorphism invariance of
amplitudes, given that all the typical manipulations with differential
operators and their integration by parts are performed in this case in a
covariant way.  In this section, we will concentrate on the simplest
cases, namely, maximally symmetric spaces, and we will give here the
renormalized expressions for a number of loop diagrams.  In these
spaces, the $\xi R \phi^2$ coupling can be reabsorbed into the mass term
and, thus, there is no need for a separate renormalization of $\xi$.

\bigskip

{\it a) The hyperboloid} $H_n$\\

$H_n$ is the maximally symmetric Euclidean space of constant negative
curvature in $n$ dimensions.  The simplest construction of maximally
symmetric spaces is made through the embedding into an
$(n+1)$-dimensional flat space, where the isometries of these spaces are
very clearly put in evidence.  Specifically for $H_n$, this
embedding is given by:
\beq{\eta_{AB}\, y^Ay^B=-(y^0)^2+y^\m y^\m=-\frac{1}{a^2}\,
,\label{2.1}}\eeq
where $y^A, A=0,...,n$ are coordinates in $\IR^{n+1}$ with metric
$\eta_{AB}=(-+...+)$, $\m=1,...,n$, and $R=-n(n-1)a^2$ is the (constant)
Ricci scalar of $H_n$. With this embedding, the metric in these spaces
is given by
\beq{ds^2=(\d_{\m\n}-\frac{a^2y_\m y_\n}{1+a^2\r^2})dy^\m
dy^\n\, ,\label{2.2}}\eeq
with $y_\m\equiv \d_{\m\n}y^\n$, and $\r^2\equiv y^\m y^\m$.
The action for a conformally coupled free scalar field is
\beq S={1\over 2}\int d^n\!y\ \sqrt{g(y)}\left(g^{\m\n}(y)\
\pa_\m\f(y)\ \pa_\n\f(y)+m^2\f (y)^2+{(n-2)\over 4(n-1)}\ R\ \f (y)^2
\right)\, .\label{2.3}\eeq
Denoting by $\sq =g^{\m\n}\nabla_\m\pa_\n\equiv -a^2\D$ the scalar
Laplace-Beltrami operator, the equation satisfied by the scalar
propagator in this theory is:
\beq{\left[-\sq +(m^2-{n(n-2)a^2\over4})\right] G_m(y)=
\frac{\d^{(n)}(y)}{\sqrt{g(y)}}\, .\label{2.4}}\eeq

In maximally symmetric spaces it is also possible to arrive at these
propagators {\it via} a purely group-theoretical description based on
highest weight representations of the isometry groups in these
spaces\cite{vil}.  For the hyperboloid $H_n$ this group is
$SO(1,n)$,
and if $L_{AB}$, $A,B=0,1,\ldots ,n$, are the corresponding generators,
then highest weight representations are labeled by the eigenvalue $\eta$
of $L_{0n}$ (and possibly other eigenvalues related to the spin of the
representation).  Unitarity bounds $\eta$ to be $0$ or $\ge (n-3)/2$ in
a continuous range.  Invariance under boosts fixes uniquely the
propagator, without reference to the free action for the corresponding
field.  However, we do not need the details of such a construction;
rather, for our purposes here we just use the fact that $\eta$ labels
the scalar propagator just as the mass does, and we shall use this
notation here for reasons which will become evident when we consider the
renormalization of different diagrams.  For now, the only difference
that entails is that the equation satisfied by the scalar propagator,
now labeled $G_\eta$, reads
\beq{[\D +\eta (\eta -n+1)]G_\eta (y) =
\frac{\d^{(n)}(y)}{a^2\sqrt{g(y)}}\, ,\label{2.5}}\eeq
where $a^2\eta (\eta -n+1)=m^2-n(n-2)a^2/4$ is the quadratic Casimir
eigenvalue for the representation $\eta$.  In order to find the
propagator explicitly, we first write down the laplacian:
\beq -a^2\D =(1+a^2\r^2)\frac{\pa^2}{\pa\r^2}+\left( \frac{(n-1)}{\r}+
na^2\r\right)\frac{\pa}{\pa\r}\, .\label{2.6}\eeq
It is possible to put this in a more workable form
by defining the reduced variable $z\equiv\sqrt{1+a^2\r^2}$, with range
$1\le z<\infty$. With this definition, Eq. (\ref{2.5}) becomes
\beq{\left[-(z^2-1)\frac{\pa^2}{\pa z^2}-
nz\frac{\pa}{\pa z}+\eta (\eta -n+1)\right]G_\eta (z) =
\frac{\d^{(n)}(y)}{a^2}\, ,\label{2.7}}\eeq
where we write $\d^{(n)}(y)$ in the original variables $y$ for
compactness, and where we have used $\sqrt{g(y)}\mid_{y=0}=1$.
It is then straightforward to find that
\beq{G_\eta (z)=(z^2-1)^{-\m/2} h_\eta (z)\, ,\label{2.8}}\eeq
where $h_\eta$ satisfies an associated Legendre equation, with solutions
given by $P_\n^\m (z)$ and $Q_\n^\m (z)$, the associated Legendre
functions, with $\n =\eta -n/2$ and $\m =(n-2)/2$ (for this and all
further references to these functions, cf. \cite{G-R}).  By examining
the short distance limit $z\rightarrow 1$ of these functions it is easy
to discard $P_\n^\m (z)$ as a solution since that leads to a propagator
which is regular at the coincident point, and we finally find that the
appropriate propagator is given by\footnote{To be sure, this argument
cannot discard a combination of both $Q_\n^\m$ and $P_\n^\m$, but the
highest-weight construction indeed does.}:
\beq G_\eta (z)={1\over 2\p^{n/2}}\left({a^2\over 2e^{i\p}
\sqrt{z^2-1}}\right)^{(n-2)/2}  Q_{\eta-n/2}^{(n-2)/2}(z)\,
.\label{2.9}\eeq
{}From now on we consider $n=4$ for simplicity, although it is clear that
analogous procedures will be possible in any number of
dimensions. The normalized propagator becomes
\beq{G_\eta (z)=-{a^2\over 4\p^2}\frac{1}{\sqrt{z^2-1}}
Q_{\eta-2}^1(z)\, ,\label{2.10}}\eeq
and it satisfies the equation
\beq{[\D +\eta (\eta -3)]G_\eta (z)={\d^{(n)}(y)\over a^2}\,
.\label{2.11}}\eeq
In particular, the massless propagator obtains for $\eta =2$ (and for
$\eta =1$, but this cannot be reached continuously from a massive
theory, and we discard it here), and the equation it satisfies is:
\beq{
[\D -2]\ G_{\eta =2}(z)=[\D -2]\ \frac{1}{4\p^2}{a^2\over z^2-1}=
{\d^{(4)}(y)\over a^2}\, .\label{2.12}}\eeq

We now proceed to find the DR identity appropriate to the diagram of
Fig. 1. As in the case of massive fields in flat space \cite{massive},
we first study these identities away from contact, and then include the
appropriate contact terms which lead, in this case, to a well-defined
flat limit.  The simplest case, the massless bubble diagram of Fig. 1,
exemplifies well the procedure: we attempt to write that amplitude as
(covariant) differential operators acting on less singular quantities,
and find the identity \beq{16\p^4 G_{\eta=2}^2(z)= \left(\frac{a^2}
{\sqrt{z^2-1}} Q_0^1(z)\right)^2= \frac{a^2}{2}[\D -2]\left[
\frac{a^2zQ_0^1(z) Q_0^0(z)}{\sqrt{z^2-1}}\right]\,\,\ ,\ \ z\not= 1\,
.\label{2.13}}\eeq The differential operator on the r.h.s. is manifestly
covariant, which implies we have preserved diffeomorphism invariance by
this procedure.  Also, for generic $\eta$, $Q_{\eta -2}^1(z)$ is
expressed in terms of hypergeometric functions which we cannot write in
terms of simple functions.  Thus, although we could have derived Eq.
(\ref{2.13}) more easily with the known explicit forms for $Q_0^1(z)$
and $Q_0^0(z)$, in general we cannot do that, and have to rely solely on
recursion relations among these functions in order to find the
appropriate DR identities.

The contact terms for this DR identity will bring in a renormalization
mass scale, and will determine the correct renormalization prescription
for this diagram at all $z\ge 1$. They are gotten by examining the flat
limit above ($z=\cosh ar$, $a\r =\sinh ar$, where $r$ is the geodesic
distance to the origin in $H_n$):
\bea {\displaystyle
16\p^4G^2_{\eta =2}(z)=}&{\displaystyle {1\over\r^4}\ \
{\buildrel a^2\to 0\over\longrightarrow}\ \ {1\over r^4}
\phantom{xxxxxxxxxxxxxxxx}}\cr
{\displaystyle {a^2Q_0^1(z)\over\sqrt{z^2-1}}=}&{\displaystyle
-{1\over\r^2}\ \
{\buildrel a^2\to 0\over\longrightarrow}\ \  -{1\over r^2}
\phantom{xxxxxxxxxxxxxx} }\cr {\displaystyle
zQ_0^0(z)=}&{\displaystyle {z\over 2}\ln
{z+1\over z-1}\,\,\,\,\, {\buildrel a^2\to 0\over
\longrightarrow }\,\,\,\, -{1\over 2}\ln a^2 r^2\phantom{xxxxx} }\cr
{\displaystyle \sq\, =} &{\displaystyle\phantom{xxx}
\pa_\r^2+{3\over\r}\pa_\r+a^2(\r^2\pa_\r^2+4\r\pa_\r )\ \
{\buildrel a^2\to 0\over\longrightarrow}\ \ \sq_{\rm flat}\  ,}
\eea
leading to:
\beq
{1\over r^4}=-{1\over 4}\, \sq_{\rm flat}\, {\ln a^2 r^2\over r^2}
\phantom{xxxx} {\rm as}\,\,\,\, a^2\to 0\, .
\eeq
The r.h.s. of this is singular as $a^2\to 0$, and we cure that by
introducing a finite local quantity to the DR identity Eq.
(\ref{2.13}): \beq
16\p^4G_{\eta =2}^2(z)=
\frac{a^2}{2}(\D -2)\left[
\frac{a^2zQ_0^1(z)Q_0^0(z)}{\sqrt{z^2-1}}\right]+\p^2\ln{M^2\over a^2}
\,\, \d^{(4)}(y)\, .\label{contact}
\eeq

This then accomplishes the renormalization of the massless bubble
diagram, with the usual DR prescription that the identity is now to be
understood for {\it all} $z\ge 1$, and $\D$ as acting through
integration by parts.  The full, renormalized 1-loop 4-point amplitude
in $\l\f^4$ theory is obtained by adding the tree-level amplitude to
$\l^2/2$ times the above one, for each of $s,t,u$ channels.  When this
is required to satisfy the appropriate renormalization group equation,
the correct value of the $\b$-function, Eq.  (\ref{betafunction}), is
obtained. Again, the simplicity of $H_n$ allowed us to circumvent
an independent $\xi$-renormalization.

At this point, this renormalized amplitude should be compared with the
general procedure outlined in the previous section.  There we exploited
the fact that, generically, the short-distance behavior is the same as
in flat space, and thus ultraviolet renormalization is achieved by
``imitating" as much as possible, given only $G$ and ${\cal D}$
generically, the renormalization of the flat bubble diagram, Eq.
(\ref{boxtrick}).  Thus, $1/x^2$ is substituted by $G$, $\ln x^2M^2$ by
$-\ln G/M^2$, and the result is Eq.  (\ref{genbox}).  That does
accomplish the renormalization of $G^2$ generically, but at the cost
of introducing an extra distribution $D$ which corrects for the
long-distance discrepancies this ``imitation" process entails.  Indeed,
we have independently checked that the same generic procedure
works here, and does lead to the correct $\b$-function.  However, Eq.
(\ref{contact}) shows clearly that in this specific case {\it we can do
better than that}, i.e., we have a specific, explicit expression valid
at all distances without complicated distributions left over.  This will
not happen in the finite temperature case, as we shall see in Sec. 3,
but does also happen for massive fields in flat space \cite{massive}.
The reason for this is that, loosely speaking, both in $H_4$ and for
massive fields, just like $Q_\n^1$ and $K_1$ (the modified Bessel
function) are the respective equivalents of $1/x$ in flat space, so are
$Q_\n^0$ and $K_0$ the respective equivalents of the flat space quantity
$\ln x^2M^2$, and thus in these settings renormalization logarithms
enter naturally while still allowing us to exploit recursion relations
among these special functions.  On the other hand, in these cases the
renormalization scale will always enter separately, determined through a
limiting procedure as above.  As a final comment, we note also that
while $M^2$ enters Eq.  (\ref{contact}) and the generic Eq.
(\ref{genbox}) in different ways, it does so in such a way that the
logarithmic mass derivative of the renormalized amplitude gives a purely
local term, which is uniquely responsible for the value of the 1-loop
$\b$-function.

We can now generalize the above to all $\ell$-loop 2-point functions of
the form $(G_{\eta =2}(z))^{\ell+1}$.  It is not difficult to verify the
following identities for the corresponding 2-, 3-, and 4-loop diagrams,
for $z\not= 1$:
\bea {\displaystyle
\left(\frac{a^2}{\sqrt{z^2-1}}
Q_0^1(z)\right)^3=}&{\displaystyle
-\frac{a^4}{16}(\D +4)(\D -2)\left[
a^2z(Q_0^1(z))^2Q_0^0(z)\right]\phantom{xxxxxxxxxxxxxxxxxxxx}}\cr
{\displaystyle
\left(\frac{a^2}{\sqrt{z^2-1}}
Q_0^1(z)\right)^4=}&{\displaystyle
\frac{a^6}{384}(\D +18)(\D +4)(\D -2)\left[
a^2z\sqrt{z^2-1}(Q_0^1(z))^3Q_0^0(z)\right]\phantom{xxxxxx}}\cr
{\displaystyle
\left(\frac{a^2}{\sqrt{z^2-1}}
Q_0^1(z)\right)^5=}&{\displaystyle \
\frac{-a^8}{18432}(\D +40)(\D +18)(\D +4)(\D -2)
\left[ a^2z(z^2-1)(Q_0^1(z))^4Q_0^0(z)\right]}\, .\label{highloop}
\eea
The first identity above, when supplemented by the appropriate contact
term, represents the renormalized setting sun amplitude in $\l\f^4$
theory.  Also in $\l\f^4$ theory, the second identity is a vacuum energy
diagram, while the third one, being made of 5-point vertices, would not
occur.  For the setting sun diagram, the same discussion as above for
the bubble diagram and its generic renormalization applies here, with
the generic renormalization given in Eq.  (\ref{settingsun}), and the
ensuing anomalous dimension for $\f$ is given by Eq.  (\ref{anomalous}).

A pattern is now seen to emerge here which is valid for all $\ell$: for
the $\ell$-loop diagram, the differential operators which appear in the
renormalization are given by:
\beq
\prod_{k=1}^\ell [ \D +k\eta (k\eta -3)]\, ,\label{thresh}
\eeq
with $\eta =2$, and the renormalizing quantity on the r.h.s. at each
loop is written as
\beq
a^2z(z^2-1)^{\ell -3\over 2}(Q_0^1(z))^\ell Q_0^0(z)\, ,
\eeq
which in fact is always the same as the 1-loop one, from $Q_0^1(z)= -1/
\sqrt{z^2-1}$, but is written thus in order to preserve the same power
of Legendre functions appearing on the l.h.s..  A comment is in order
regarding the differential operators renormalizing these diagrams.  In
the flat massive case \cite{massive}, it was noted that these operators
indicate the particle production thresholds for the corresponding
diagrams, thus $(\sq -4m^2)$ for the bubble, $(\sq -9m^2)$ for the
setting sun diagram, etc..  Here, a new but related phenomenon is seen
to occur: if we consider the propagators here to carry the
representation $\eta =2$ (of the isometry group $SO(1,4)$ of $H_4$),
then the differential operators in Eq.  (\ref{thresh}) will count the
number of representations propagating along the diagram minus one.  In
curved spaces we cannot attach this the same meaning as in flat space,
i.e., that this is an indication of the analiticity structure of the
respective $p$-space amplitudes, but we nonetheless find this curious
phenomenon worthy of mention.  For conciseness, we do not present here
the contact terms appropriate for the renormalization of these higher
loop diagrams, but they are straightforward to find in the same fashion
as for the bubble diagram.

We can now move on to the renormalization of massive diagrams. They are
somewhat more difficult to work out, and we present here the results
only for the bubble and setting sun diagrams. In what follows, we need
to use two recursion relations for associated Legendre functions
\cite{G-R}: \bea {\displaystyle
{dQ_\n^\m(z)\over dz}=}&{\displaystyle
\m {z\over z^2-1}Q_\n^\m(z)+{1\over\sqrt{z^2-1}}
Q_\n^{\m +1}(z)\phantom{xxxxxxxxxxxxxxx}}\cr {\displaystyle
Q_\n^{\m +2}(z)=}&{\displaystyle
-2(\m +1){z\over\sqrt{z^2-1}}Q_\n^{\m +1}(z)+ (\n -\m)
(\n +\m +1)Q_\n^\m(z)}\, .\label{recursion}
\eea

For the bubble diagram, we find after some manipulations:
\beq
\left(\frac{a^2}{\sqrt{z^2-1}}
Q_{\eta -2}^1(z)\right)^2=\frac{a^2}{2}\left(\D +4{m^2\over
a^2}-2\right) \left[
\frac{a^2zQ_{\eta -2}^1(z)Q_{\eta -2}^0(z)}{\sqrt{z^2-1}}\right]+
a^2m^2(Q_{\eta -2}^0(z))^2\, ,\label{massbubbl}
\eeq
where $m^2=(\eta -1)(\eta -2)a^2$ (compare Eq.(\ref{2.4}) with
(\ref{2.5}) for $n=4$ to find this).  We note that, again, in the
differential operator appearing here, although we cannot speak of an
analiticity structure for the corresponding $p$-space amplitude, we also
see a ``production threshold" for two particles of mass $m$ appearing
besides the curvature effect in the term $-2$ analyzed previously.
Finally, to obtain the renormalized amplitude corresponding to this
diagram, we need the appropriate contact terms.  We find these in a
two-step process: first, we take the limit $a^2\to 0$ while holding
$m^2$ fixed.  This is done with the help of the following limit
\cite{A-S}:
\beq
\lim_{\n\to\infty}\,\,\,\n^{-\m}e^{-\m\p i}Q_\n^\m(\cosh
{\xi\over\n})=K_\m(\xi )\,\, ,\,\,\,\,\, \xi >0,\label{bessel}
\eeq
where $K_\m$ are modified Bessel functions.  By using this in the DR
identity above, with $z=\cosh ar$ and $\n^2\sim m^2/a^2$, we find
precisely the DR identity for massive scalars on flat space
\cite{massive}:
\beq
\left( {mK_1(mr)\over r}\right)^2={1\over 2}\, (\sq -4m^2)\, {mK_0(mr)
K_1(mr)\over r}\,\, .
\eeq
So far, no contact terms need be added. Now, we determine contact
terms by the massless limit of the above equation. We simply borrow the
result from \cite{massive}, and finally write for the
renormalized amplitude:
\beq
\left(\frac{a^2}{\sqrt{z^2-1}}
Q_{\eta -2}^1(z)\right)^2=
\frac{a^2}{2}\left(\D +4{m^2\over a^2}-2\right)
\left[ \frac{a^2zQ_{\eta -2}^1(z)Q_{\eta -2}^0(z)}{\sqrt{z^2-1}}\right]+
a^2m^2(Q_{\eta -2}^0(z))^2+\p^2\ln (16M^2/\g^2 m^2)\,\,\d
^{(4)}(y)\,\, ,\label{mbubblcont}
\eeq
where $\g$ is the Euler-Mascheroni constant.  Naturally, we could also
have taken the direct $a^2\to 0,m^2\to 0$ limit to determine these
contact terms, and the result would be identical to what we found for
the massless case, Eq.  (\ref{contact}).  These two procedures differ by
a choice of scheme or, more specifically, by the finite contact term \\
$\p^2\ln [(\eta -1)(\eta -2)\g^2/16]\d^{(4)}(y)$.

We now present the DR identity that renormalizes the massive setting sun
diagram. It represents somewhat more effort than previous cases, and we
find:

\bea & {\displaystyle -16 \left(\frac{a^2}{\sqrt{z^2-1}} Q_{\eta
-2}^1(z)\right)^3=}\cr
&{\displaystyle
a^4 \left[ \D +9 {m^2\over a^2}+4\right] \left[ \D
+{m^2\over a^2} -2 \right] \left( a^2 z Q^0_{\eta-2 }(z)( Q^1_{\eta-2
}(z))^2+m^2 z ( Q^0_{\eta-2 }(z))^3\right)
-32 a^4 m^2
\sqrt{z^2-1} \left( Q^1_{\eta-2 }(z)\right)^3}\cr &{\displaystyle
-8 m^2a^2( a^2+8m^2)
\sqrt{z^2-1}\left(Q^0_{\eta-2 }(z)\right)^2 Q^1_{\eta-2 }(z)
-28
m^2 a^2\left( a^2 z Q^0_{\eta-2 }(z)( Q^1_{\eta-2 }(z))^2+m^2 z (
Q^0_{\eta-2 }(z))^3\right)}\label{settingsunhn}\eea

By using the limit Eq. (\ref{bessel}), the above identity again goes
very smoothly in the $a^2\to 0$, $m^2$ fixed limit into the
corresponding amplitude in \cite{massive}. Borrowing the massless limit
of that expression, like for the setting sun above, determines the
contact terms appropriate for this DR identity. The result above stands
as a success of DR, insofar as it produces the fully renormalized 2-loop
2-point function in a compact and explicit way, something which had not
been achieved previously.
\bigskip

{\it b) The sphere} $S_n$\\

$S_n$ is the maximally symmetric space of constant positive
curvature in $n$ dimensions.  Because it is a more familiar space, and
because most formulas here would be essentially redundant with formulas
from the previous section on the hyperboloid, we simply give here a more
condensed summary of our results.  Essentially all the geometry formulas
from last section can be adjusted to spheres by taking $a^2\to -a^2$.
The reduced variable to be used here, rather than $z$, is
$x\equiv\pm\sqrt{1-a^2\r^2}$, with range $-1\le x\le 1$, where the
positive(negative) sign holds for the upper(lower) hemisphere.  The
massive propagator is
\beq G_m (x)={1\over 2\p^{n/2}}\left({a^2\over 2e^{i\p}
\sqrt{1-x^2}}\right)^{(n-2)/2}  Q_{\n}^{(n-2)/2}(x)\,
,\label{b2.9}\eeq
where now $\n (\n +1)=-m^2/a^2$, and it satisfies the equation
\beq{\left[-(1-x^2)\frac{\pa^2}{\pa x^2}+
nx\frac{\pa}{\pa x}+\left({m^2\over a^2}+{n(n-2)\over
4}\right)\right]G_m (x) = \frac{\d^{(n)}(y)}{a^2}\,
.\label{b2.7}}\eeq
For the massless propagator in $n=4$ in particular, we have:
\beq{G_{m=0} (x)=-{a^2\over 4\p^2}\frac{1}{\sqrt{1-x^2}}
Q_{0}^1(x)\, ,\label{b2.10}}\eeq
satisfying
\beq [\D +2]\ G_{m=0} (x) =
\frac{\d^{(4)}(y)}{a^2}\, .\label{b2.5}\eeq
The renormalization of $\ell$-loop 2-point functions, analogously to
Eqs. (\ref{2.13}) and (\ref{highloop}), works as follows:
\bea {\displaystyle
\left(\frac{a^2}{\sqrt{1-x^2}}
Q_0^1(x)\right)^2=}&{\displaystyle
\frac{a^2}{2}(\D +2)\left[
{a^2xQ_0^1(x)Q_0^0(x)\over\sqrt{1-x^2}}\right]}
\phantom{xxxxxxxxxxxxxxxxxxxxxxxxx}\cr{\displaystyle
\left(\frac{a^2}{\sqrt{1-x^2}}
Q_0^1(x)\right)^3=}&{\displaystyle
-\frac{a^4}{16}(\D -4)(\D +2)\left[
a^2x(Q_0^1(x))^2Q_0^0(x)\right]}\phantom{xxxxxxxxxxxxxxxxxx}\cr
{\displaystyle
\left(\frac{a^2}{\sqrt{1-x^2}}
Q_0^1(x)\right)^4=}&{\displaystyle
\frac{a^6}{384}(\D -18)(\D -4)(\D +2)\left[
a^2x\sqrt{1-x^2}(Q_0^1(x))^3Q_0^0(x)\right]}\phantom{xxxxxx}\cr
{\displaystyle
\left(\frac{a^2}{\sqrt{1-x^2}}
Q_0^1(x)\right)^5=}&{\displaystyle \,\,
\frac{-a^8}{18432}(\D -40)(\D -18)(\D -4)(\D +2)
\left[ a^2x(1-x^2)(Q_0^1(x))^4Q_0^0(x)\right]}\, .\label{bhighloop}
\eea

The same pattern has emerged here as previously, with the only
difference that all signs are reversed.  Again, we do not work out the
contact terms for these DR identities, as the procedure is identical to
what was done previously.  Also, one can verify that the first two DR
identities above, which lead to the renormalization of the bubble and
setting sun diagrams, respectively, again lead to the correct values for
the $\b$-function and anomalous dimension in $\l\f^4$ theory at lowest
nontrivial order.

With the hindsight of our work on the hyperboloid, it becomes
straightforward to check the renormalization identities for the massive
bubble and setting sun amplitudes on the sphere. They are given,
respectively, by:
\beq
\left(\frac{a^2}{\sqrt{1-x^2}}
Q_{\n }^1(x)\right)^2=\frac{a^2}{2}\left(\D +4{m^2\over
a^2}+2\right) \left[
\frac{a^2xQ_{\n }^1(x)Q_{\n }^0(x)}{\sqrt{1-x^2}}\right]-
a^2m^2(Q_{\n }^0(x))^2\, ,\label{massbubblsph}
\eeq
and
\bea & {\displaystyle - 16 \left(\frac{a^2}{\sqrt{1-x^2}} Q_{\n }^1
(x)\right)^3=}\cr &{\displaystyle a^4\left[\D +9{m^2\over a^2}-4\right]
\left[\D +{m^2\over a^2}+2\right]\Bigl( a^2xQ^0_\n (x)(Q^1_\n (x))^2
+m^2 x ( Q^0_{\n }(x))^3\Bigr) +32 a^4 m^2 \sqrt{1-x^2} \left(
Q^1_{\n }(x)\right)^3}\cr &{\displaystyle -8 a^2m^2( a^2-8m^2)
\sqrt{1-x^2}\left(Q^0_{\n }(x)\right)^2 Q^1_{\n }(x) +28 m^2
a^2\Bigl( a^2 x Q^0_{\n }(x)( Q^1_{\n }(x))^2+m^2 x ( Q^0_{\n
}(x))^3\Bigr)\ ,}\label{settingsunsph}\eea
and appropriate contact terms are again gotten by a straightforward
massless and flat limit above.\bigskip

\section{renormalization at finite temperature}\bigskip

Although quantum field theory is usually defined at zero temperature, it
is also possible to consider it at arbitrary nonzero temperatures
\cite{kapusta,bernard,jackiw}, and finite temperature
effects are relevant, for instance, in studying phase transitions in the
early universe \cite{linde} or the thermal structure of QCD
\cite{qcd}.  This implementation of temperature in quantum
field theory is essentially made in one of two ways: either one proceeds
in real Minskowskian time, or goes into Wick-rotated, compactified
Euclidean time.  The former suffers from some ambiguities \cite{AVBD},
while the latter is a more appropriate setting for our approach, and is
the one we shall follow here.

Again, to illustrate our general procedure, we work here with massless
$\lambda \f^4$ theory.  It is known that finite temperatures induce
masses through renormalization, and thus a consistent, renormalizable
scalar model in this context {\it must} be massive.  This phenomenon
will be seen very clearly in our treatment from the non-closure of the
renormalization group equations for the 2-point function.  We choose,
nonetheless, to start with a massless theory here in order to illustrate
our method in a simple way.  Sure enough, the renormalization group
equation will not be verified due to the lack of a bare mass to cancel a
temperature-induced mass counterterm appearing at two loops.  We then
switch on a bare mass and treat it as a correction to the massless case
in order to get its appropriate renormalization.  The spirit of our
approach remains, to show that in any theory one is able to isolate
singular terms and correct them using differential identities.

In imaginary time formalism, time is rotated into Euclidean time $\tau$
and compactified to a cylinder of perimeter
\beq{\beta={1\over k T}\ ,}\eeq
where $k$ stands for the Boltzmann constant.  The momentum space
propagator is
\beq {1\over w_n-\vec{p}\,{}^2}\qquad,\qquad w_n=i\,{2\pi n\over
\beta} \ ,\label{propft}\eeq
where the Euclidean zeroth component of the momentum vector takes only
discrete values.  In coordinate space this propagator can be written
using the method of images as
\beq{G(x,\tau;\beta)={1\over 4\pi^2} \sum_{n=-\infty}^{ \infty} \
{1\over x^2+(\tau+n \beta)^2} \ ,\label{3.2}}\eeq
where $x=\sqrt{\vec{x}^2}$.  This expression can be resummed, or,
alternatively, the Fourier transform of the momentum propagator can be
computed to yield
\bea {\displaystyle G(
x,\tau;\beta)}&{\displaystyle ={1\over 4\pi\beta x}\,{\sinh
{2\pi\over \beta} x \over \cosh {2\pi\over \beta}x -\cos {2\pi\over \beta}
\tau}}\cr & {\displaystyle = {\nu\over 8\pi^2 x}{\sinh \nu
x\over \cosh \nu x-\cos \nu \tau}}\ , \label{3.3} \eea
where we have rewritten the last line using $\nu\equiv
{2\pi\over\beta}=2\pi k T$.  Note that this compact form of the
propagator is manifestly periodic in Euclidean time and decreases at
long distances as $1/x$, which is the characteristic decay in three
dimensions.  This is a hint at the effective reduction from four
dimensions to three at high temperatures.

The simplest diagram is a pure tadpole. Its departure from
$\beta=\infty$ can easily be computed in coordinate space using the
first form of the propagator, Eq. (\ref{3.2}). The computation
is reduced to
\beq{\lim_{(x,\tau)\to (0,0)}\left(G(x,\tau;\beta)-
G(x,\tau;\beta=\infty)\right)=\sum_{n\not= 0} {1\over 4\pi^2}
{1\over (n \beta)^2}={1\over 12\beta^2}\ .\label{3.4}}\eeq
This result agrees with the standard literature \cite{kapusta,jackiw}.
It can also be obtained through a simple expansion in Eq.  (\ref{3.3}).
Nontrivial renormalization first appears when trying to compute the
vertex function at 1-loop order.  We need to renormalize $G^2$, the
square of the propagator, which presents a logarithmic singularity.  We
proceed as in Sec. 1, by noting that for $x\ne 0$
\beq{G^2(x)={1\over 16 \pi^2}\, \sq \left( G(x) \ln G(x)/M^2\right) +
{\n^2\over 16\pi^2} G(x) - {\n\over 128 \pi^4
 x^3}{\left( \n x \cosh \n x - \sinh \n x\right)^2\over
\sinh \n x \left( \cosh \n x-\cos \n \tau\right) }\ ,\label{3.5}}\eeq
where $G(x)$ is short-hand for the propagator and
the laplacian operator is
\beq{
\sq= \partial_\tau\partial_\tau+{1\over x^2}\partial_x  x^2\partial_x
\ .\label{3.6}}\eeq
The renormalization of this diagram has been carried out according to
the generic analysis presented in Sec. 1. The singularity has been
corrected in the laplacian term, and the last two terms of the r.h.s.
above correspond to the expected left-over distribution, $D$, which
encodes extra long-distance information of the theory.  One can easily
check that this distribution $D$ has a leading singularity of the type
$( x^2+\tau^2)^{-1}$, which is Fourier-transformable.

Now it is easy to compute the first coefficient of the $\beta $
function, using the renormalization group equation for the 4-point
amputated vertex function.  At this order, there is no contribution
coming from the anomalous dimension of the field $\phi$ , thus Eq.
(\ref{betafunction}) is readily obtained and we explicitly see that, to
this order, the short-distance renormalization of the theory has not
been affected by the compactification of Euclidean time.  This
corroborates the fact that we are dealing with a scheme-independent term
of the $\b$-function.  To compute this term of the $\b$-function,
only the coefficient of the logarithm matters, so we do not need the
actual form of $D(x)$, provided it is a well-behaved distribution.  Yet,
the point is that we have explicitly obtained the finite
parts of the diagram without much effort.

The 2-loop correction to the vertex function consists of two diagrams,
Figs. 2,3.  Fig. 2, as explained in Sec. 1, is just a convolution of the
1-loop diagram and only promotes logarithms to their squares, as
dictated by the first coefficient of the $\b$-function, giving no
contribution to the second coefficient of the $\b$-function.  The other
one is computed exactly as in Sec. 1, that is, in a first step we
renormalize the inner divergence through Eq.  (\ref{firststep}), with
${\cal D}=-\sq$, and $D$ given in Eq.  (\ref{3.5}) above.  Identical
manipulations lead to Eqs.  (\ref{generictwoloop},\ref{dprime}), where
now $f(x)=0$.  Again, the relative coefficient $-2$ between $\ln^2$ and
$\ln$ above is dictated by short-distance behavior, and thus it is fixed
and universal.  Thus, collecting all the pieces, we finally obtain the
renormalized expression
\bea &{\displaystyle -{\lambda^3\over 3}
\Biggl\{
G(x-y)G(x)D(x)+{1\over
16\pi^2}{\partial \over
\partial y^{\mu}}\left[ G(x-y)G(x){\fletxeta \over \partial
y^{\mu}}\ln {G(y)\over M^2}\right] }
\cr&{\displaystyle -{1\over
2}{1\over \left( 16\pi^2\right)^2}\delta^{(4)} (x-y)\,\sq\left[
G(y)\left( \ln^2{G(y)\over M^2}-2\ln
{G(y)\over M^2}\right) \right] -{1\over
16\pi^2}\delta^{(4)}
(x-y)D(y)\Biggr\}
} \eea
Where, again, we note that we have easily obtained
explicitly the finite terms.

We now turn to the self-energy correction depicted in Fig. 4, and again
use the general recipe given in Sec. 1. Without undue effort, one finds
\beq{G^3(x)={1\over 512\pi^4}\left( \,\sq +2\n^2\right) \sq
G(x)\ln {G(x)\over M^2}+R(x)\ ,\label{settingsunft}}\eeq
where
\bea {\displaystyle R(x)=}&{\displaystyle -{\n \over 4096\pi^6}\sq
\left( {1\over x^3}{\left(\n x \cosh \n x-\sinh \n x\right)^2\over \sinh
\n x\left( \cosh \n x-\cos \n \tau \right) }\right) +{\n^4\over
256\pi^4}G(x)}\cr&{\displaystyle -{\n^2\over 2048\pi^6}{1\over
x^4}{\left( \n x \cosh \n x-\sinh \n x\right)^2\over \left( \cosh \n
x-\cos \n \tau \right)^2}{2\sinh \n x+\n x\cosh \n x-\n x \cos \n \tau
\over \sinh \n x}} \ .\eea
This has the same short-distance behavior as the propagator, and is
Fourier-transformable.

As explained in Sec. 1, the appearance of a term linear in the laplacian
(with coefficient $2\n^2$) has to do with subleading singularities.  In
this case, it reflects the fact that the thermal bath screens the
interaction.  This will generate the inconsistency we are expecting:
the renormalized 2-loop 2-point function will be
\beq{\Gamma^{(2)}=-\sq \delta^{(4)} (x)-{\lambda^2\over
6}\left( {1\over 512\pi^4}\left( \sq +2\n^2\right) \sq G(x)\ln
{G(x)\over M^2}+R(x)\right) \label{selfen}}\ ,\eeq
and to check whether it satisfies the appropriate RG equation we
compute
\beq M{\pa\over\pa M} \G^{(2)}={\l^2\over 6} \left({1\over 16
\p^2}\right)^2\left(\sq\d^{(4)} (x)+2\n^2\delta^{(4)} (x)\right)\
.\eeq
The first term leads to the expected value of the anomalous dimension,
Eq. (\ref{anomalous}), while the second term represents a mass
renormalization. That means that, as mentioned previously,
renormalizability requires a massive theory {\it ab initio}.

It is clear that masses modify the long-distance behavior of the
propagator while preserving the $1/x^2$ singularity at the origin.
Since renormalization consists in curing short-distance singularities,
the possibility is open to treat masses as perturbations.  More
precisely, the exact massive propagator at finite temperature can be
found to be
\beq G(x,\t ;\beta;m)\ =\ {\nu\over 8\pi^2 x}\ \sum^\infty_{n=-\infty}
e^{-i n \nu \tau- x\sqrt{\nu^2 n^2+m^2}}\ .\label{ massprop}\eeq
Expanding around the massless case, we find
\beq G(x,\t ;\b ;m)\ =\ {\n \over 8 \p^2 x} \left( {\sinh \n
x \over \cosh \n x - \cos \n \t } - m x + {m^2 x \over 2 \n}
\ln \left[ \n^2 \left( x^2 + \t^2 \right) \right]
\right) + ... \label{massiveprop} \eeq
We are now ready to use this modified expression of the propagator to go
through the renormalization procedure in all previous graphs. It is
readily noticed that the 1- and 2-loop  coefficient of the
$\beta$-function and the 2-loop coefficient of the anomalous dimension
of the scalar field are not modified by the presence of masses. The
first and relevant appearance of a mass counterterm takes place in the
setting sun  diagram, as explained above.
 The expression for that diagram corresponds to Eq.
(\ref{selfen}) with the addition of the first
massive correction \beq   \left({\l \over 16 \p^2}\right)^2
 \left\{ \n m \, \sq G(x) \ln {G(x) \over M^2}
+ {m^2 \over 4} \, \sq \left[
G(x) \left(
\ln^2 \left[ \n^2 \left( x^2 + \t^2 \right) \right] - 2 \ln {G(x) \over
M^2} \right) \right] \right\} +
... \eeq
The complete 2-point function then verifies the renormalization group
equation
 \beq \left(
 M{\partial\over \partial M}+\beta {\partial\over \partial \lambda}
+\gamma_m m^2{\partial\over\partial m^2}-2 \gamma\right)
\Gamma^{(2)}(x) =0 \label{csftm}\eeq
with
\beq \g_m = \left({\lambda\over 16 \pi^2}\right)^2 \left( {1 \over 3}
\left(
{\n \over m} \right)^2 - 2 {\n \over m } + {7 \over 6} \right) +
O(\l^3) \label{massrenormalization} \eeq
which is a scheme-dependent coefficient.

\bigskip
\section{conclusions} \bigskip

The aim of the present paper has been to exploit the idea that the
renormalization process amounts to replacing short-distance
singularities by distributions and, more importantly, that this project
can be carried out for any background a field theory might be defined
on.  Explicit examples on constant curvature and thermal backgrounds
have been worked out.

We can draw several lessons from this investigation.  In the finite
temperature case, the absence of a compact form for the massive
propagator in position space has forced us to treat masses in a
perturbative way.  This is a drawback, though renormalization group
constants are obtained probably more easily than with other methods.  On
the other hand, constant curvature backgrounds are surprisingly suited
for differential renormalization.  Explicit closed expressions for
2-loop amplitudes are obtained with ease.

Further applications of this renormalization procedure are manifold, as
for instance the extension to QED, along the lines of \cite{QED}, or
computations of the effective potential on curved backgrounds.  Its very
essence, based on the coordinate space propagator, makes it an extremely
simple method to renormalize a field theory.

\bigskip
\section*{Acknowledgments}\bigskip

We thank K. Kirsten and S. Odintsov for comments on the
manuscript.

\end{document}